\documentclass{aa}
\usepackage{amsmath}
\usepackage{graphicx}
\usepackage{media9}
\usepackage{xcolor}
\usepackage{natbib}
\usepackage{dirtytalk}
\usepackage{hyperref}
\usepackage{array,tabularx}
\usepackage{multimedia}
\usepackage[switch]{lineno}

\definecolor{myblue}{RGB}{51,90,140}
\newcommand{\myblue}{\textcolor{myblue}}

\begin{document}
\let\oldpageref\pageref
\renewcommand{\pageref}{\oldpageref*}

\title{\Large \textbf{Spectral softening in core-collapse supernova remnant expanding inside wind-blown bubble
}}

\author{\textbf{\myblue{Samata Das}}\inst{1,2}\thanks{\myblue{samata.das@desy.de}}
\and{\textbf{\myblue{Robert Brose}}}\inst{3}
\and{\textbf{\myblue{Dominique M.-A. Meyer}}}\inst{2}
\and{\textbf{\myblue{Martin Pohl}}}\inst{1,2}
\and{\textbf{\myblue{Iurii Sushch}}}\inst{4,5}
\and{\textbf{\myblue{Pavlo Plotko}}}\inst{1}}

\institute{{Deutsches Elektronen-Synchrotron DESY, Platanenallee 6, 15738 Zeuthen, Germany} \and
{Institute of Physics and Astronomy, University of Potsdam, 14476 Potsdam, Germany} \and
{Dublin Institute for Advanced Studies, 31 Fitzwilliam Place, Dublin 2, Ireland} \and
{Centre for Space Research, North-West University, 2520 Potchefstroom, South Africa} \and
{Astronomical Observatory of Ivan Franko National University of L’viv, vul. Kyryla i Methodia, 8, L’viv 79005, Ukraine}}

\date{Received 25 November, 2021/ Accepted 02 March, 2022}

\abstract{
\noindent{\textbf{\myblue{Context.}} Galactic cosmic rays are widely assumed to arise from diffusive shock acceleration, specifically at shocks in supernova remnants (SNRs). These shocks expand in a complex environment, particularly in the core-collapse scenario as these SNRs evolve inside the wind-blown bubbles created by their progenitor stars. The cosmic rays (CRs) at core-collapse SNRs may carry spectral signatures of that complexity.}\newline
{\textbf{\myblue{Aims.}} We study particle acceleration in the core-collapse SNR of a progenitor with initial mass $60M_{\odot}$ and realistic stellar evolution. The SNR shock interacts with discontinuities inside the wind-blown bubble and generates several transmitted and reflected shocks. We analyse their impact on particle spectra and the resulting emission from the remnant.}\newline
{\textbf{\myblue{Methods.}} 
To model the particle acceleration at the forward shock of SNR expanding inside a wind bubble,  we have initially simulated the evolution of the pre-supernova circumstellar medium by solving the hydrodynamic equations for the entire lifetime of the progenitor star. As the large-scale magnetic field, we have considered parameterised circumstellar magnetic field with passive field transport.} Then, the hydrodynamic equations for the evolution of SNR inside the pre-supernova circumstellar medium have been solved simultaneously with the transport equation for cosmic rays in test-particle approximation and with the induction equation for the magnetohydrodynamics (MHD) in 1-D spherical symmetry.\\
{\textbf{\myblue{Results.}} The evolution of core-collapse SNRs inside complex wind-blown bubbles modifies the spectra of both the particles and their emission, on account of several factors including density fluctuations, temperature variations,} and the magnetic field configuration. We have found softer particle spectra with spectral indices close to 2.5 during shock propagation inside the shocked wind, and this softness persists at later evolutionary stages. Further, our calculated total production spectrum released into the interstellar medium demonstrates spectral consistency at high energy with the galactic CRs injection spectrum, required in propagation models. The magnetic field structure effectively influences the emission morphology of SNR as it governs the transportation of particles and the synchrotron emissivity. There rarely is a full  correspondence of the intensity morphology in the radio, X-ray, and gamma-ray bands.}
    
\keywords{Supernova Remnants - Bubbles - Cosmic Rays}
\titlerunning{Particle acceleration in SNR}
\authorrunning{S. Das et al.}
\maketitle


\section{Introduction}

Supernova Remnants (SNRs) are major sources of galactic cosmic rays (CRs) below the \say{knee} energy ($\approx 10^{15} \, \mathrm{eV}$) \myblue{\citep{1934PNAS...20..259B,  2013A&ARv..21...70B}}. The acceleration mechanism which can accelerate CRs to this high energy is thought to be the widely studied Diffusive Shock Acceleration (DSA) process  \myblue{\citep{1949PhRv...75.1169F, 1978MNRAS.182..147B, 1983RPPh...46..973D}} and its non-linear modification \myblue{\citep{1997ApJ...487..197E}}. According to DSA, the spectra of particles accelerated at SNR shock should follow a power law in energy with the spectral index 2 with an exponential cut-off at the maximum achievable energy, limited spatially by the size, temporally by the age of SNRs, as well as by radiative energy losses and adiabatic cooling. In recent years, observations in the TeV band (HESS, VERITAS, MAGIC, HAWC, and LHAASO) and in the GeV band (AGILE, and \textit{Fermi}-LAT)  have been collecting a significant amount of data regarding SNRs which provide crucial insight as well as constraints for theoretical models. Spectral measurements of gamma-ray emission from, e.g., IC443 
\myblue{\citep{2009ApJ...698L.133A}}, Cas A \myblue{\citep{2010ApJ...710L..92A}}, SN 1006 \myblue{\citep{2010A&A...516A..62A}}, Tycho’s SNR \myblue{\citep{2011ApJ...730L..20A}}, and W44 \myblue{\citep{2011NatCo...2..194M, 2014A&A...565A..74C}} indicate a considerable softening compared to the expected power-law index $s=2$, which may be modelled in different ways. For example, diffusive re-acceleration of galactic CRs has been proposed to explain the spectral shape of W44 \myblue{\citep{2016A&A...595A..58C}}, but was found implausible in other studies on account of the large thickness of radiative shocks and the paucity of Galactic cosmic rays to be re-accelerated \myblue{\citep{2020A&A...634A..59B,2020MNRAS.497.3581D}}. Other options include re-acceleration in fast-mode turbulence downstream of the forward shock \myblue{\citep{2015A&A...574A..43P,2020A&A...639A.124W}}, fast motion of downstream turbulence \myblue{\citep{2020ApJ...905....2C}}, and inefficient particle confinement in the vicinity of the SNR caused by the attenuation or weak driving of Alfvén waves \myblue{\citep{2011NatCo...2..194M, 2019MNRAS.490.4317C, 2020A&A...634A..59B}}.

{The CR acceleration at SNR shocks depends on the type of SNRs and the hydrodynamic and magnetic-field structure of its environment. Specifically, the morphology of core-collapse SNRs (e.g. \myblue{\citet{1989ApJ...344..332C, 1989A&A...215..347C, 2005ApJ...630..892D,   2007ApJ...667..226D,2021MNRAS.502.5340M}}), carries the signature of the type of progenitor stars, for instance "ear"-like morphology for Luminous Blue Variable (LBVs) progenitors \myblue{\citep{2021MNRAS.502..176C, 2021arXiv210801951U}}}.   
 CR acceleration at shocks propagating through stellar wind was discussed in  \myblue{\citep{1988ApJ...333L..65V, 1989A&A...215..399B, 2000A&A...357..283B}} considering Bohm diffusion of energetic particles. The changes in particle spectra in core-collapse scenario with Red Super Giant (RSG) and Wolf-Rayet (WR) star progenitors were investigated in \myblue{\citet{2013A&A...552A.102T}} for simplified flow profiles.
Most recently, the effects of the circumstellar magnetic field on electron spectra and subsequent non-thermal emissions were studied in \myblue{\citet{2021arXiv211106946S}}, focusing on the impact originated during the transition of SNR forward shock from the free wind to the shocked wind region of the wind bubble. In both of these studies, the complete hydrodynamic evolution of the circumstellar medium (CSM) during stellar evolution has 
not been taken into account, however a realistic representation of the CSM at the pre-supernova stage can potentially impose  better constrains than the ones already demonstrated. \myblue{\citet{2021arXiv211106946S}} also investigated the impact on the synchrotron cooling of a parametrised post-shock magnetic field amplification.
Besides possibly amplifying the magnetic field, resonant and non-resonant streaming instabilities determine the spectrum of turbulence, and hence the diffusion coefficient as well as the maximally attainable particle energy at any point in time during the evolution of the remnant \myblue{\citep{2020A&A...634A..59B}}. For this study a detailed consideration of turbulent magnetic field is out of scope. A forthcoming study including magnetic field amplification with realistic hydrodynamics shall explore the additional impact of CR-driven instabilities on particle acceleration as well as radiation from the remnant.
\par In this paper, we investigate the spectral modification for CRs accelerated at the forward shock of an SNR, as it evolves through the different regions of the wind bubble, simulated using an evolutionary track for Zero Age Main Sequence (ZAMS) mass $60M_{\sun}$.  
We present the imprint of interactions of SNR forward shock with multiple shocks and contact discontinuities inside the wind bubble on the particle spectra and demonstrate that the obtained CR spectra are softer than predicted for strong shocks.
Additionally, we illustrate the effects of the circumstellar magnetic field along with the hydrodynamics. To study the SNR with $60\,M_{\odot}$ progenitor is interesting as emissions from the SNR with this massive progenitor star can be predicted theoretically, although it is not a frequent event \myblue{\citep{2014ApJ...795..170J}}. Furthermore, a $60M_{\sun}$ star is thought to evolve through luminous blue variable (LBV) phase instead of RSG phase, but ends its life as a WR star, all of which leave their imprint in the morphology of the ambient medium and hence on the particle spectra.

\section{Numerical methods}
We introduce the reader to the numerical methods used in this study. The diffusive shock acceleration (DSA) at SNR forward shock has been modelled in test-particle approximation. The necessary constituents for this modelling are a hydrodynamic description of the CSM structure, a large-scale magnetic field profile, a prescription for diffusion, and finally the solution for CR transport equation. We have numerically solved the particle acceleration and hydrodynamics, respectively, with RATPaC (Radiation Acceleration Transport Parallel Code) \myblue{\citep{2012APh....35..300T,2013A&A...552A.102T, 2020A&A...634A..59B, 2018A&A...618A.155S}} and the PLUTO code \myblue{\citep{2007ApJS..170..228M,vaidya_apj_865_2018}}. 
\par This section begins by presenting the hydrodynamics of pre-supernova circumstellar medium in which we have inserted a supernova explosion. Then, the structure of magnetic field followed by the method for calculating the particle acceleration are described.

\subsection{Hydrodynamics}
\label{subsec:2.1}
The Euler hydrodynamic equations including an energy source/sink term can be expressed as (considering the magnetic field too weak to become dynamically important):
\begin{linenomath*}
\begin{equation}\label{eq:1}
   { \frac{\partial}{\partial t} \,   
\begin{pmatrix}
\rho \\
\bf m\\
E
\end{pmatrix}}
+ \nabla
 \,
\begin{pmatrix}
\rho\bf u \\
\textbf{m} \textbf {u} + P \bf I\\
(E+P)\bf u
\end{pmatrix}
^T =   
\,
\begin{pmatrix}
0 \\
0\\
S
\end{pmatrix}
\end{equation}
\end{linenomath*}
\begin{linenomath*}
\begin{equation}\label{eq:2}
    \frac {\rho \textbf {u}^2}{2}+\frac{P}{\gamma-1} = E; \quad \gamma = \frac{5}{3}
\end{equation}
\end{linenomath*}

where $\rho$, \textbf {u},\textbf{ m}, P, E, $S$ are the mass density, velocity, momentum density, thermal pressure, the total
energy density, and source/sink term, respectively. \textbf{I} is the unit tensor.  

\subsubsection{Construction of CSM at pre-supernova stage}\label{subsec:2.1.1}
To simulate the wind bubble created by a non-rotating $60M_{\odot}$ star at solar metallicity (Z = 0.014) from ZAMS to pre-supernova stage, we have performed a hydrodynamic simulation with PLUTO in 1-D spherical symmetry. For the simulation, the computational domain [$O,R_{\mathrm{max}}$] with origin $O$ 
and $R_{\mathrm{max}}=150\, \mathrm{parsec}$ has been discretised into 50000 equally spaced grid points. The interstellar medium is assumed to have a constant number density, $n_\mathrm{ISM}=1 \mathrm{atom\, cm^{-3}}$. To initialise the simulation, a radially symmetric spherical supersonic stellar wind has been injected into a small spherical region of radius $0.06$~pc at the origin, using the stellar evolutionary track for $60M_{\odot}$ ZAMS described in \myblue{\citet{2014A&A...564A..30G}}. The wind density, $\rho_{\mathrm{wind}}$, can be written as:
\begin{linenomath*}
\begin{equation}\label{eq:3}
\rho_{\mathrm{wind}} = \frac{\dot{M(t)}}{4\pi r^{2} u_{\mathrm{wind}}(t)}\ ,
\end{equation}
\end{linenomath*}
where r is the radial coordinate, and $\dot{M}$ and $u_{\mathrm{wind}}$ represent the time-dependent mass-loss rate and the wind velocity, respectively, that have been taken from \myblue{\citet{2014A&A...564A..30G}}. 
To model the evolution of the wind bubble, Equations \eqref{eq:1} and \eqref{eq:2} have been integrated with a second-order Runge-Kutta method as well as using the Harten-Lax-Van Leer approximate Riemann Solver (hll) and finite volume methodology. Further, optically-thin cooling and radiative heating have been included through the source/sink term, $S = \Phi(T,\rho)$, using the cooling and heating laws described in \myblue{\citep[][Sec. 2.3]{2020MNRAS.493.3548M}}. The time steps for the simulation have been constrained using the standard Courant-Friedrich-Levy (CFL) condition, initialised as $C_{\mathrm{cfl}}=0.1$. 

The stellar evolution has been followed from zero age to the pre-supernova phase at $3.95\, \mathrm {million\, years}$. The state of the CSM at this time is the initial state of the SNR simulation. Therefore, this model is an 1D equivalent 
of the 2D simulation in the static-star scenario presented in \myblue{\citet{2020MNRAS.493.3548M}}. The stellar wind parameters at the post-main sequence stages are illustrated in \myblue{\citep[][Sec. 2.4]{2020MNRAS.493.3548M}}. 
\subsubsection{Modelling of supernova ejecta profile}\label{subsec:2.1.2}

The density distribution of supernova ejecta is modelled as constant, $\rho_\mathrm{c}$, up to $r_\mathrm{c}$, followed by a power law to the ejecta radius, $R_{\mathrm{ej}}$:
\begin{linenomath*}
\begin{equation}\label{eq:4}
     \begin{split}
        \rho_{\mathrm{ej}}(r) = \begin{cases}\rho_\mathrm{c},  \qquad\qquad\qquad\quad r \leq r_c\\[.5em]
                \rho_\mathrm{c}\left(\frac{r}{r_\mathrm{c}}\right)^{-{n}} \quad\qquad r_\mathrm{c} < r \leq R_{\mathrm{ej}}\ ,
                \end{cases}
    \end{split}
\end{equation}
\end{linenomath*}
where $\mathrm{n} = 9$ is conventionally used for core-collapse explosion. The velocity profile for the ejecta reflects homologous expansion:
\begin{linenomath*}
\begin{equation}\label{eq:5}
       u_\mathrm{ej} =\frac{r}{T_\mathrm{SN}}\ ,
\end{equation}
\end{linenomath*}
where $T_\mathrm{SN}= 3\ \mathrm{years}$ is the start time of the hydrodynamic simulation. The initial ejecta temperature is set to $10^4$K.
\par The expressions for $r_\mathrm{c}$ and $\rho_\mathrm{c}$ can be written as a function of the ejecta mass, $M_\mathrm{ej}$, and explosion energy, $E_\mathrm{ej}$,
\begin{linenomath*}
\begin{equation}\label{eq:6}
       r_\mathrm{c} = \left(\frac{10E_{\mathrm{ej}}}{3M_{\mathrm{ej}}}\,\frac{n-5}{n-3}\,\frac{n-3x^{3-n}}{n-5x^{5-n}}\right)^{1/2}T_{\mathrm{SN}}
\end{equation}
\end{linenomath*}
\begin{linenomath*}
\begin{equation}\label{eq:7}
      \rho_\mathrm{c} = \frac{M_{\mathrm{ej}}}{4\pi r_\mathrm{c}^3}\frac{3(n-3)}{n^2}\left(n-3x^{3-n}\right)^{-1}\ ,
\end{equation}
\end{linenomath*}
In our simulation, $R_{\mathrm{ej}} = xr_\mathrm{c}$ and $x = 2.5$, $E_\mathrm{ej}=10^{51}\ \mathrm{erg}$, and $M_\mathrm{ej}= 11.75\,\mathrm{M_{\sun}}$\footnote{\myblue{$M_\mathrm{ej} = M_{\star} (60M_{\sun}) - \int_{t_{\mathrm{t_{ZAMS}}}}^{t_{\mathrm{preSN}}} \dot{M}(t)\mbox dt- M_{\mathrm{Compact Object}}(1.4M_{\sun})$}}, respectively.

\subsubsection{Hydrodynamic modelling to study SNR shock evolution}\label{subsec.2.2.1}
\begin{figure}
\centering
\includegraphics[width=\columnwidth]{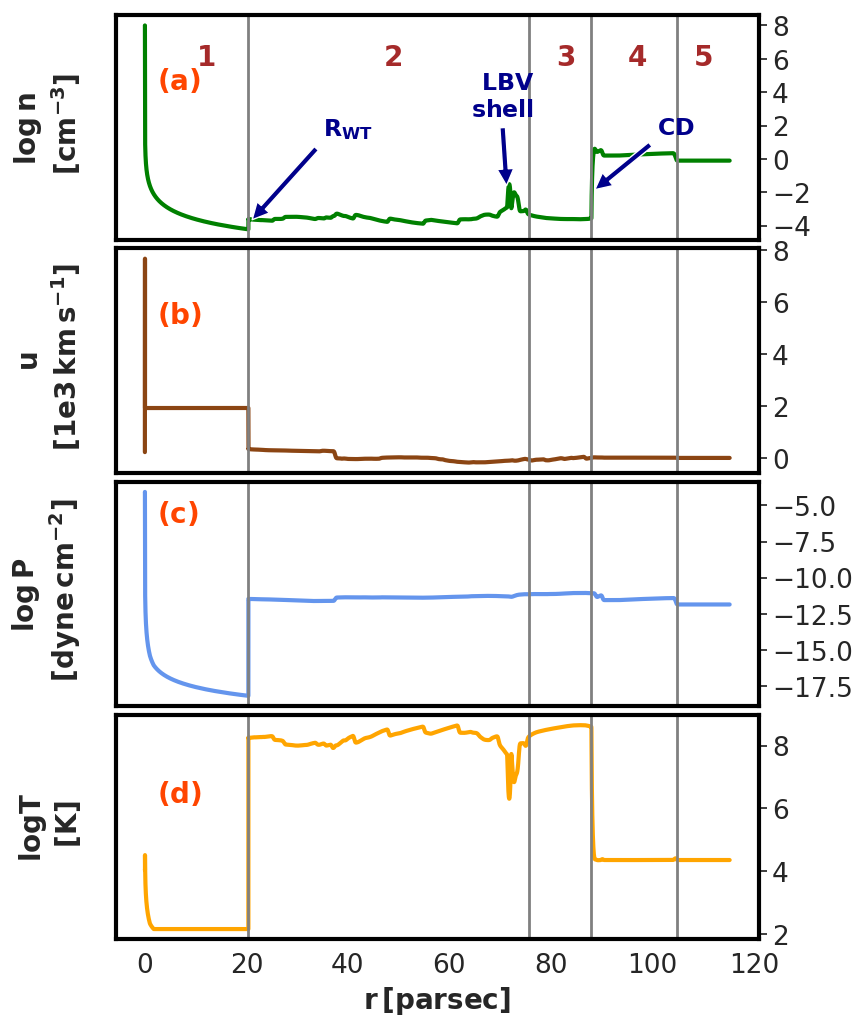}
\caption{\textbf{Profiles of the number density (n) (panel (a)), the flow speed (u) (panel (b)), the thermal pressure (P) (panel (c)), and the temperature (T) (panel (d)), right after the supernova explosion.}
Vertical grey lines mark the boundary of the supernova ejecta (up to 0.023 parsec), the free stellar wind (region 1), the shocked LBV and WR wind (region 2), the shocked wind from the O and B phases (region 3), the shocked interstellar medium (ISM) (region 4), and the ambient ISM (region 5). $\mathrm{R_{WT}}$ is the radius of wind termination shock, LBV shell denotes the dense shell created by interaction between LBV wind and WR wind, and CD represents the contact discontinuity between shocked wind and shocked ISM}
\label{fig: Figure 1}
\end{figure}

To initiate the supernova explosion, 
 the 
supernova ejecta profile has been inserted in, and interpolated with the pre-calculated pre-supernova CSM profile, illustrated in Fig.\ref{fig: Figure 1}. Then, to model the evolution of the SNR, Equations \eqref{eq:1} and \eqref{eq:2} have been solved considering the local source  as zero ($S=0$) using a Harten-Lax-Van Leer approximate Riemann Solver that restores with the middle contact discontinuity (hllc), finite-volume methodology, and a second-order Runge-Kutta method. The numerical simulation with the PLUTO code has been performed in 1-D spherical symmetry with 262144 uniform grid cells with $R_\mathrm{max}=112\ \mathrm{pc}$ to provide a spatial resolution of about $0.0004$~pc.\footnote{\myblue{We cut the grid of the pre-supernova CSM because we follow the SNR shock only to the shocked ISM. Additionally, we increased by interpolation the grid resolution from 50000 cells to 262144 cells.}}

\subsection{Magnetic field}
\subsubsection{Field profile}\label{subsec:2.2.2}

To acquire the large-scale magnetic field profile for the entire lifetime of SNR, we have solved the induction equation for ideal magnetohydrodynamics (MHD) following \myblue{\citet{2013A&A...552A.102T}}. This method mimics MHD for negligible magnetic pressure.
The structure of the CSM magnetic field is quite intricate specifically in the presence of the different evolutionary stages of the massive star \myblue{\citep[see for a rotating O star][] {2020JPhCS1620a2012M}}. Therefore, modelling the CSM magnetic field with MHD simulation for the entire life of the $60M_{\odot}$ star is out of scope for this paper, but for simplicity, we can parametrise the CSM magnetic field using background information about the stellar magnetic field. 
\par The wind of a rotating star carries off mass and magnetic field. In the presence of a weak magnetic field, the flow speed is as for a non-magnetic wind, and the magnetic field becomes frozen-in \myblue{\citep{1991IAUS..143..289C}}. Gauss' law ($\mathbf{\nabla\cdot B} = 0$) gives the expression for radial field,
\begin{linenomath*}
\begin{equation}\label{eq:9}
    B _\mathrm{r} = B_{\star}\frac{R_{\star}^2}{r^2}.
\end{equation}
\end{linenomath*}
For a rotating star, the toroidal field in the equatorial plane of rotation \myblue{\citep{1998ApJ...505..910I}} can be written as \myblue{\citep{1999IAUS..193..325G,1994ApJ...421..225C}},
\begin{linenomath*}
    \begin{equation}\label{eq:10}
       B_{\phi} 
       =  B_{\star} \frac{u_\mathrm{rot}R_{\star}}{u_\mathrm{wind}r} \qquad\qquad r >> R_{\star}\ ,
    \end{equation}
    \end{linenomath*}
where $B_{\star}$ and $R_{\star}$ are the stellar surface magnetic field and radius, respectively, $u_\mathrm{rot}$ and $u_\mathrm{wind}$ represent the surface rotational velocity in the equatorial plane and the radial wind speed, respectively. The toroidal field will be strongly dominant except for very close to the stellar surface. The radial field can be expected to provide an impact only during the first days of the SNR evolution discussed in \myblue{\citet{2021ApJ...922....7I}}, and hence out of scope of this paper. 
\par Using the wind profiles of a non-rotating $60 M_{\odot}$ star and the rotation of the WR star, the pre-supernova stage of a $60M_{\odot}$ star, we parametrised the circumstellar magnetic field, $B_\mathrm{CSM}$. 
The surface magnetic field and stellar radius have been set to $1000\ \mathrm{G}$ and $6\,R_{\odot}$, respectively, following \myblue{\citet{2007ARA&A..45..177C}}. The wind speed and surface rotational velocity have been approximated to 2000$\ \mathrm{km}\,\mathrm{s}^{-1}$ and 100$\ \mathrm{km}\,\mathrm{s}^{-1}$, respectively, following \myblue{\citet{1996ApJ...459..671I,2010ApJ...716..929C}}. The magnetic field is compressed by a factor 4 at the wind termination shock, as is the density. For simplicity, we have considered a constant field strength in the shocked wind, as a significantly more realistic model would require MHD simulations and assumptions about the magnetic field at the launch point of the stellar wind throughout the entire evolution of the progenitor star.
Therefore, the magnetic field in the regions marked in Fig. \ref{fig: Figure 1} can be approximated as,
\begin{equation}\label{eq:11}
B_\mathrm{CSM} = \begin{cases}(0.33\ \mathrm{\mu G})\frac{R_{\mathrm{WT}}}{r} & \mathrm{region}\, 1,\, \mathrm{beyond}\,0.023 \,\mathrm{parsec}\\[.5em]
       1.32\ \mathrm{\mu G} \qquad &  \mathrm{regions}\,2 \,\& \,3\\[.5em]
       15.6\ \mathrm{\mu G}  \qquad &  \mathrm{region}\, 4\\[.5em]
       4.5\ \mathrm{\mu G} \qquad &   \mathrm{region}\, 5\ .
\end{cases}
\end{equation}

\par For region 4, magnetic field ($B_\mathrm{shell}$) has been calculated as in \myblue{\citet{2015A&A...584A..49V}},

\begin{equation}\label{eq:12}
     B_\mathrm{shell} =\frac{R^2B_\mathrm{ISM}}{R^2 - (R-d)^2}\ ,
\end{equation}
where $B_\mathrm{ISM}$ is the interstellar medium (ISM) magnetic field, and R and d are the outer radius and the thickness of region 4, respectively. The magnetic field in the free wind is too weak to make the magnetic pressure dynamically important. The strength of $B_\mathrm{ISM}$ is chosen to provide super-Alfvénic motion of the shell in region 4 into the ISM, meaning we allow for an outer shock in the ISM.
\par For the initial magnetic field in the supernova ejecta, $B_\mathrm{ej}(r) \propto 1/r^2$ satisfies $\mathbf{\nabla \cdot B_\mathrm{ej}} = 0$ and $\mathbf{\nabla}\times(\mathbf{u}_{\mathbf{ej}}\times \mathbf{B}_{\mathbf{ej}})=0$ for both the radial and the toroidal field component. The normalisation is chosen to provide a volume-averaged magnetic field of $30\,\mathrm{G}$ when the SNR radius is $10^{15}$ cm (explained elaborately in \myblue{\citep[][Sec.3]{2013A&A...552A.102T}}), 

\begin{equation}\label{eq:13}
        B_{\mathrm{ej}}(r,t_0) = \left(10\,\sqrt{3}\ \mathrm{G}\right)\, \left(\frac{(10^{15}\ \mathrm{cm})\,R_{\mathrm{ej}}}{r\,R_\mathrm{sh}(t_0)}\right)^2  \quad  r \leq R_{\mathrm{ej}}\ .
\end{equation}

where $R_\mathrm{sh}$, $t_{0}$ are the SNR shock radius and starting time of simulation, respectively. 

The subsequent evolution is that of frozen-in magnetic field and given by the induction equation in 1D spherical symmetry \myblue{\citep{2013A&A...552A.102T}},
\begin{equation}\label{eq:14}
    \frac{\partial \mathbf{B}}{\partial t} = \mathbf{\nabla}\times(\mathbf{u}\times \mathbf{B})\ .
\end{equation}

\subsubsection{Diffusion coefficient}
The diffusion coefficient directly influences the acceleration time scale and the maximum attainable energy of the particles \myblue{\citep{1983A&A...125..249L, 2010MNRAS.406.2633S}}. The spatial diffusion coefficient can be expressed as,
\begin{linenomath*}
    \begin{equation}\label{eq:15}
        D = \zeta D_0 \left(\frac{pc}{10\mathrm{GeV}}\right)^\alpha\left(\frac{B}{3\mu G}\right)^{-\alpha}\ ,
    \end{equation}
    \end{linenomath*}
where $\zeta$ is a free scaling parameter and $D_0$ is the diffusion coefficient for the nominal momentum and magnetic-field strength, either $D_\mathrm{B}$ for Bohm diffusion with $\alpha = 1$ or $D_\mathrm{G}=10^{29}\ \mathrm{cm^2\mbox s^{-1}}$ with $\alpha=1/3$ for Galactic diffusion, respectively. We have applied ten times the Bohm diffusion coefficient ($\zeta D_0 = 10D_\mathrm{B}$) in the entire region downstream of the SNR forward shock, as well as galactic diffusion coefficient ($\zeta D_0 = D_\mathrm{G}$) in the far-upstream region starting from 2$R_{\mathrm{sh}}$ \myblue{\citep{2012A&A...541A.153T}}, and a connecting exponential profile between these two regions. 
\par A more realistic approach would be solving the transport equation of magnetic turbulence at least for the resonant CR streaming instability. A forthcoming study will include the magnetic field fluctuations through the diffusion coefficient prescribed in \myblue{\citet{2016A&A...593A..20B}}.

\subsection{Particle acceleration}
\label{subsec:2.3}
The time-dependent transport equation for the differential number density of CRs, $N(p)$, can be expressed as
\begin{linenomath*}
 \begin{equation}\label{eq:17}
    \frac{\partial N}{\partial t} = \nabla (D\nabla N - \mathbf{u}N) - \frac{\partial }{\partial p}\left(\dot pN - \frac{\nabla\cdot\mathbf{u}}{3}Np\right) +Q
 \end{equation}
 \end{linenomath*}
where $D$ is the spatial diffusion coefficient, 
$\dot p$ corresponds to  energy loss rate (synchrotron losses and inverse Compton losses for electrons), 
$\mathbf{u}$ refers to the plasma velocity, and $Q$ represents the source term. 
\par This transport equation has been solved in test-particle approximation and for spherical symmetry with RATPaC, applying implicit finite-difference algorithms implemented in the FiPy package \myblue{\citep{2009CSE....11c...6G}}. In our simulation the cosmic-ray pressure has always remained below $10\%$ of the shock ram pressure \myblue{\citep{2010ApJ...721..886K}}. We have used a shock-centred coordinate system, $x=r/R_{\mathrm{sh}}$, where $R_{\mathrm{sh}}$ is the shock radius. Additionally, we have transformed the radial coordinate as $(x-1) = (x^*-1)^3$ to get a better spatial resolution near the shock, $\Delta r/R_\mathrm{sh}\approx 10^{-6}$. This choice also provides a grid extent to $65R_{\mathrm{sh}}$ to track the particles escaped from the vicinity of the shock but still inside the far upstream region. 
\subsection{Injection of particles} 
The source term in the transport equation is defined by,
\label{subsubsec:2.3.1}
\begin{linenomath*}
 \begin{equation}\label{eq:18}
  Q = \eta n_\mathrm{u} (V_{\mathrm{sh}}-u_\mathrm{u})\delta(R-R_{\mathrm{sh}})\delta(p-p_{\mathrm{inj}})\ ,
 \end{equation}
 \end{linenomath*}
where $\eta$ is the injection efficiency, and $n_\mathrm{u}$ and $u_\mathrm{u}$ are the upstream plasma number density and velocity, respectively, $V_{\mathrm{sh}}$ and $R_{\mathrm{sh}}$ are the shock velocity and radius, respectively, and $p_{\mathrm{inj}}$ represents the momentum of injected particles. Following \myblue{\citet{2005MNRAS.361..907B}}, the injection momentum is defined as a multiple of thermal momentum, $p_{\mathrm{inj}} = \xi p_{\mathrm{th}}=\xi\sqrt{2m k_BT_d}$. The injection efficiency is
\begin{linenomath*}
 \begin{equation}\label{eq:19}
  \eta = \frac{4}{3\pi^{1/2}}(R_\mathrm{sub}-1)\xi^3\exp{(-\xi^2)}\ ,
 \end{equation}
 \end{linenomath*}
where $R_\mathrm{sub}$ represents the compression ratio of the sub-shock. The momentum of the injected particles should be significantly larger than the thermal momentum of downstream particles to participate in shock acceleration. Although \myblue{\citet{2016ApJ...827L..29S}} demonstrated that CR-feedback has important effect on driving the galactic outflows in ISM when CRs escape the SNR at late evolutionary stage, CR-feedback is out of scope for this paper. Therefore, even though $\xi=4.24$ has been found appropriate for, e.g., SN1006 \myblue{\citep[][Appendix A]{2021A&A...654A.139B}}, we have taken $\xi = 4.4$, as for smaller values of $\xi$ the test-particle approximation would not be valid when the shock passes through the dense LBV shell and through the shocked ISM. The coupled equations for the hydrodynamic evolution of the SNR, the evolution of the large-scale magnetic field, and the transport of CRs have been solved simultaneously. The CR transport equation and the induction equation for the magnetic field can be solved with a time-step of $1$~year, but the standard CFL condition limits the time-step of the hydrodynamic simulation to $10^{-3}-10^{-4}$ years.

\section{Results}
\subsection{Shock parameters}
\label{subsec:3.1}
\begin{figure}
\centering
\includegraphics[width=9.0cm]{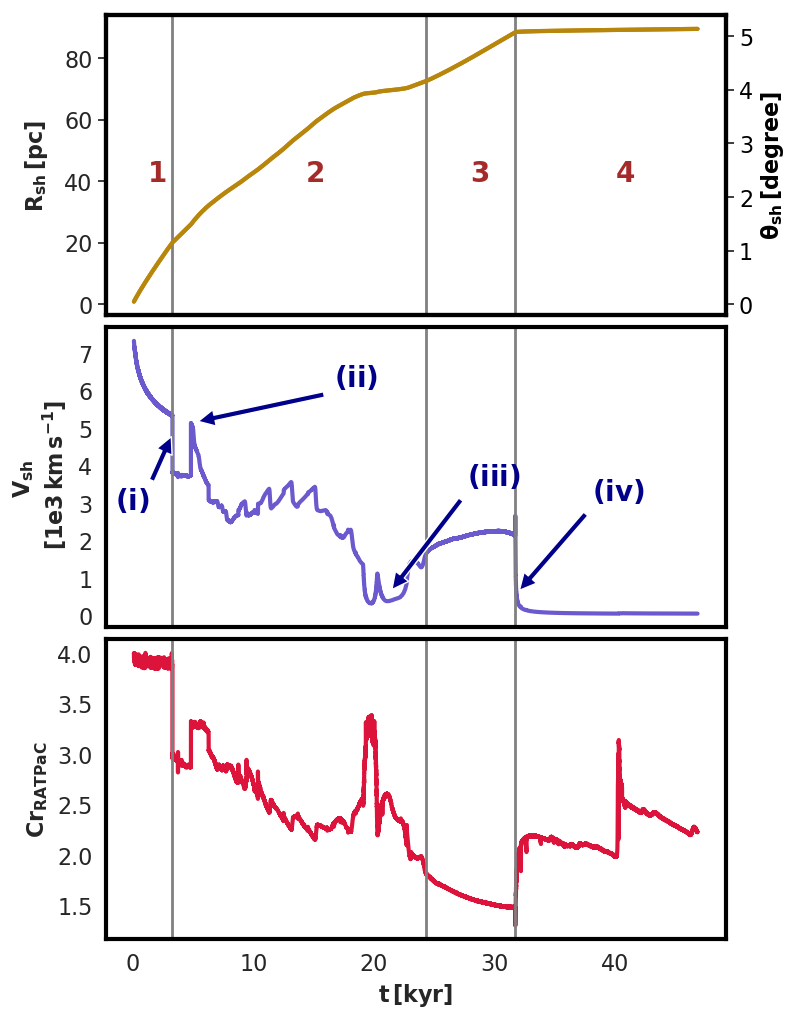}
\caption{\textbf{Behaviour of forward-shock (FS) parameters}: radius ($R_\mathrm{sh}$), velocity ($V_\mathrm{sh}$), and sub-shock compression ratio ($\mathrm{Cr_{RATPaC}}$). In the upper panel, we also provide the angular scale, $\theta_{sh}$, for a distance of $1000\,\mathrm{parsec}$. \textbf{(i)-(iv) mark interactions of the FS with different discontinuities}, namely (i) the wind termination shock, (ii) another outgoing shock, (iii) the LBV shell, and (iv) the CD.}
\label{fig: Figure 2}
\end{figure}

The evolution of the SNR with ZAMS mass $60M_{\sun}$ progenitor has been studied for 46,000 years, until the SNR forward shock (FS) starts expanding inside the shocked ISM region, and the sonic Mach number of the FS falls below 2. Fig. \ref{fig: Figure 2} illustrates the time evolution of the shock radius and velocity, as well as the sub-shock compression ratio, as the SNR FS propagates through the various regions of the wind bubble depicted in Fig. \ref{fig: Figure 1}. In the free stellar wind, the shock velocity gradually decreases from approximately $\mathrm{7300\,km\, s^{-1}}$ to $\mathrm{5300\,km\, s^{-1}}$. After about 3300 years, the FS interacts with the wind termination shock, transits to the denser shocked wind, and the shock speed plummets to $\mathrm{1500\,km\, s^{-1}}$. Then, after nearly 4830 years, the FS velocity rises steeply by $\mathrm{1400\,km\,s^{-1}}$ as a consequence of a tail-on collision between the FS and the reflection off the contact discontinuity between FS and reverse shock (RS) of the reflected shock produced during the interaction between the FS and the wind termination shock. After that, the shock velocity fluctuates a lot, on account of interactions between the FS and various weaker discontinuities in the shocked wind. 
Between 19,000 years and 23,000 years, the FS passes through the LBV shell, followed by a slight rise in shock velocity after 24000 years during the passage through the low-density shocked wind of the O and B phase of the progenitor. Finally, after 32,000 years, the FS interacts with the CD of the CSM, and the shock speed sharply falls from approximately $\mathrm{2000\,km\,s^{-1}}$ to $\mathrm{46\,km\,s^{-1}}$. 

In the free stellar wind, the shock is strong and sub-shock compression ratio is close to 4. Fig. \ref{fig: Figure 2} demonstrates that the compression ratio falls to 2.9, as soon as the FS enters the hot shocked stellar wind, and it finally becomes 1.5 right before the FS/CD interaction at 32,000 years. The variations in the sub-shock compression ratio simply reflect the change in sonic Mach number of the FS, $\mathrm{M_{s}}$. Tests verify that the numerically derived value $\mathrm{Cr_{RATPaC}} = \mathrm{v_{u}}/\mathrm{v_{d}}$, with $\mathrm{v_{u}}$ and $\mathrm{v_{d}}$ as the upstream and downstream flow speed in the FS rest frame, conforms well with the theoretical value based on the upstream temperature and the shock speed, $\mathrm{Cr_{Theoretical}}=(1+\gamma)\mathrm{M_s}^2/((\gamma-1)\mathrm{M_s}^2+2)$, where $\gamma = 5/3$. The shocked wind is hot enough to reduce the sonic Mach number to single-digit numbers.

\subsection{Particle spectra}
\label{subsec:3.2}
\begin{figure}
\centering
\includegraphics[width=9cm]{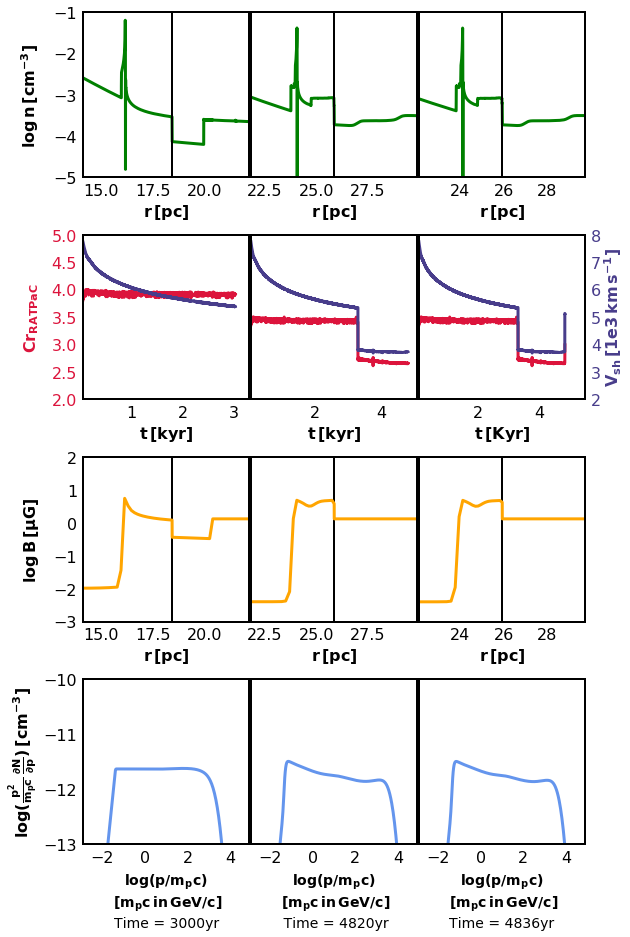}
\caption{\textbf{Proton spectra volume-averaged downstream of the FS at early times}:  For context, we also provide the gas number density, $n$, as a function of radius. The second row displays the compression ratio ($\mathrm{Cr}_\mathrm{RATPaC}$) and the shock speed ($V_\mathrm{sh}$) up to the specific age, the third row depicts the magnetic-field profile ($B$), and the fourth row illustrates the proton spectra. The vertical lines in the first and the third row mark the FS position.}
\label{fig: Figure 4}
\end{figure}
\begin{figure}
\centering
\includegraphics[width=9cm]{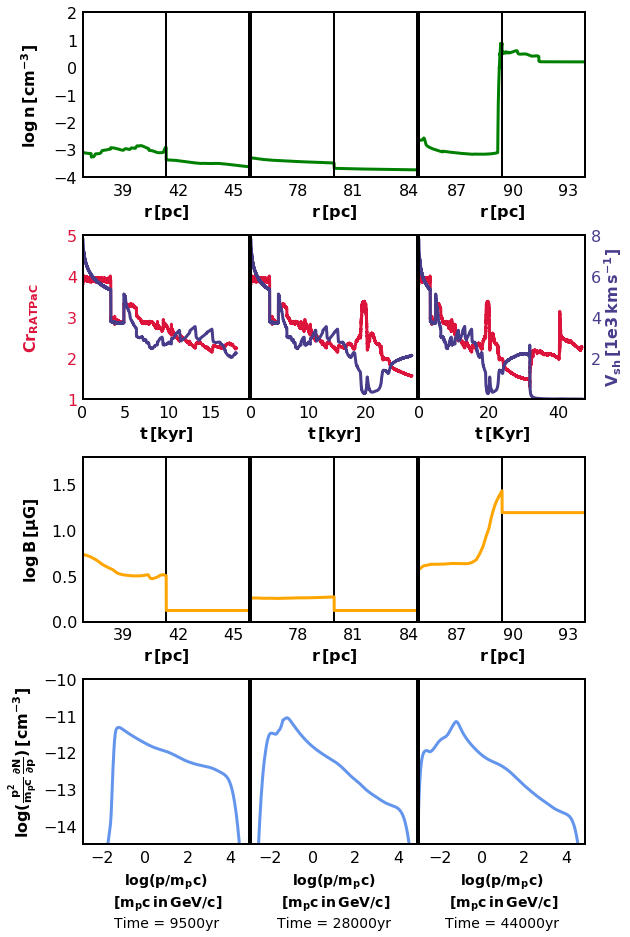}
\caption{\textbf{Proton spectra volume-averaged downstream of the FS as in Fig. \ref{fig: Figure 4}, but for later times.}}
\label{fig: Figure 5}
\end{figure}
To evaluate proton and electron spectra at times characteristic for FS propagation through the different regions of the wind bubble, we show their volume averages for the entire region downstream of the FS. The large-scale transported magnetic field described in subsection \ref{subsec:2.2.2}, shapes in particular the electron spectra through the synchrotron energy losses. Below a few tens of TeV, their importance relative to inverse-Compton (IC) losses (cosmic microwave background (CMB) photons only) scales with the energy density ratio of magnetic field and CMB photons, $ \mathrm{U_{B}} /\mathrm{U_\mathrm{CMB}}$. In our modelled magnetic field scenario, the magnetic field inside the RS is as weak as $0.01\,\mu G$, which renders inverse Compton losses dominant in the deep interior of the SNR.

\vspace{5pt}
\par \textbf{FS in the free wind:} At 3000 years the FS still propagates through the free wind and is about to interact with the wind termination shock that is located at a radius of about 20 parsec. The first and third rows of first column of Fig. \ref{fig: Figure 4} show the profiles of the gas number density, n, and the magnetic field ($B$) respectively. The magnetic field, $\mathrm{B}$, has its peak strength at the contact discontinuity between FS and RS at a radius of approximately $16$ parsec, and it becomes moderately weaker toward the FS. At this time, the proton spectrum reflects the $E^{-2}$ expected for test-particle DSA
for strong shock, and further the maximum achievable proton energy reaches $5\,\mathrm{TeV}$.
\vspace{5pt}
\par\textbf{FS in the shocked wind:} At 4820 years, the FS propagates through the shocked wind. Shown in the second column of Fig. \ref{fig: Figure 4}, the proton spectrum starts to become softer than the standard power law, $E^{-2}$, as a consequence of lower sonic Mach number of the FS in this region. Also, the spectra display convex curvature. Detailed inspection of the spatial distribution of particles reveals that the numerous transmitted and reflected shocks between the FS and the RS, that were originally spawned by the interaction between the FS and the wind termination shock, provide re-acceleration of the cosmic rays that are primarily produced at the FS of the SNR.
Any time a shock hits a discontinuity, it breaks into a transmitted shock and a reflected shock, and after a few interactions many shocks are generated. Each of them can accelerate particles with a certain spectral index and a specific maximum energy. At high energies the contribution with the hardest spectrum should dominate \myblue{\citep{1972ApJ...174..253B,2001A&A...377.1056B}}, provided the energy can be reached, and so we see a complex superposition of the acceleration yield of many shocks.
Additionally, the magnetic field downstream of the FS is very weak, and hence the diffusion coefficient is large. Therefore, sufficiently highly energetic particles can deeply penetrate the downstream region and are able to interact with some of the reflected shocks. This produces a weak but noticeable spectral break above $\mathrm{10\,GeV}$. The exact form of the spectral break may depend on the details of the magnetic-field structure, implying it could be slightly different for a full MHD model or the inclusion of a sub-grid turbulence model. This spectral break is also visible in the electron spectra at the same time, that we show in Fig. \ref{fig: Figure 7}. 
Furthermore, an outgoing shock emerging from the interaction between a reflected shock and the contact discontinuity is about to collide with the FS. Only a few years later, at 4836 years, both the shock speed and the sub-shock compression ratio increase sharply. The third column of Fig. \ref{fig: Figure 4} shows the spectrum and the related parameters just after the shock-shock tail-on collision, and no visible change is observed in the spectrum. It appears that here shock merging does not immediately change the spectral shape of the volume-averaged proton spectrum, which also have been expected from the calculations shown in Appendix {\ref{Appendix A}}. However, this interaction will increase the acceleration rate, and therefore, the acceleration efficiency of the FS, which eventually leads to a higher maximum energy of the particles, also discussed in the \myblue{\citep[][Sec. 4.2]{2021arXiv211106946S}}. 
This finding is in line with recent analyses of the spectral effects of shock-shock collisions \myblue{\citep{2020MNRAS.494.3166V}}.
Generally speaking, the time period, during which particles can see both shocks, is shorter than the acceleration time.
\par
\begin{figure}
\centering
\includegraphics[width=9cm, height=6cm]{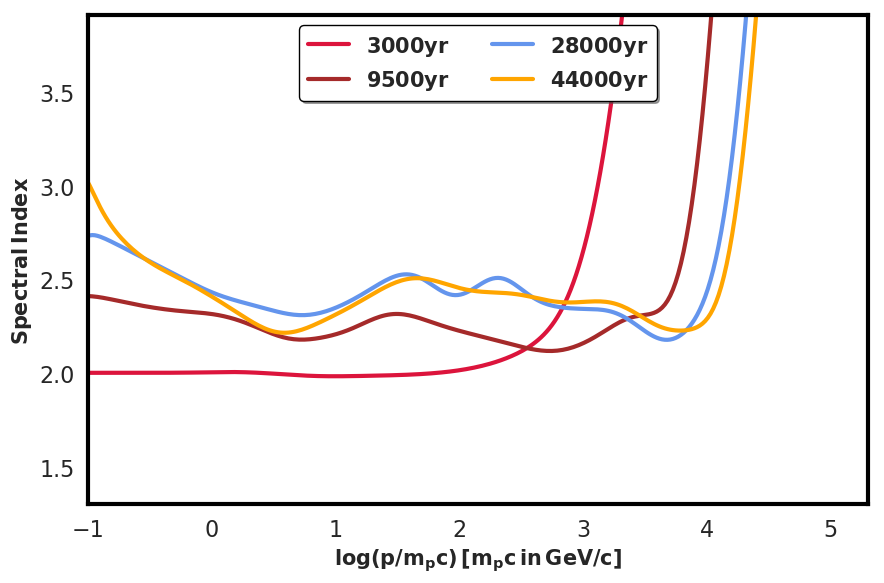}
\caption{\textbf{Variation of the spectral index for protons at different ages with momentum of SNR}.}.
\label{fig: Figure 6}
\end{figure}

\begin{figure}
\centering
\includegraphics[width=9cm, height=9cm]{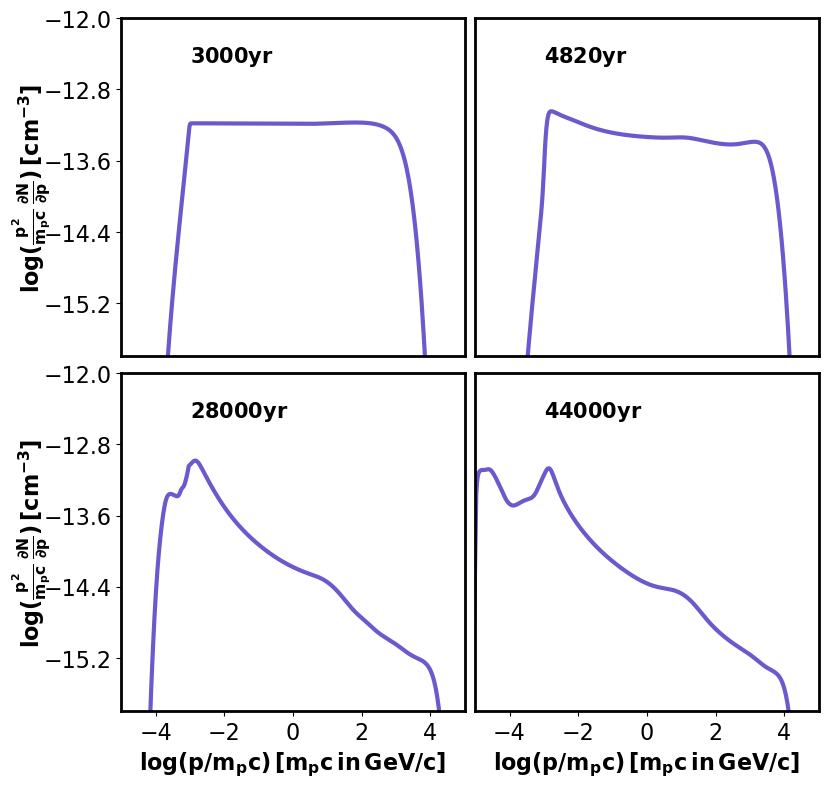}
\caption{\textbf{Electron spectra at different ages of the SNR,} volume-averaged downstream of the FS.}
\label{fig: Figure 7}
\end{figure}

 The first column of Fig. \ref{fig: Figure 5} illustrates the state after 9500 years. The FS has already gone through several interactions with different discontinuities inside the shocked stellar wind. At this time, the proton spectra are generally soft with index around $2.3$ with moderate variations between the GeV band and about $10$~TeV energies. The running spectral index at this time is also shown in Fig. \ref{fig: Figure 6}. Later, after 28000 years, the FS has entered the shocked wind from the O and B phases of the progenitor, after crossing the LBV shell. During the passage through LBV shell, the FS has encountered a relatively dense and cold material. Consequently, the injection rate of low-energy particles into DSA is high and the injection momentum is low on account of the low shock speed. The spectra in the second column of Fig. \ref{fig: Figure 5} show the pile-up at very low energy that results from this amplified injection. Here, the low-energy pile-up around $10$~MeV reflects the simplified injection model we have used. 
The softness of the spectrum above a few tens of MeV is generic though and fully arises from the small compression ratio of the FS.  The spectral index for protons reaches approximately $2.5$ at energies beyond $\mathrm{10\,GeV}$, shown in Fig. \ref{fig: Figure 6}. 

\vspace{5pt}
\textbf{FS in the shocked ISM:} The shocked ISM is $10^{4}$ times denser than the shocked O, B wind. Therefore, the collision between the FS and the wind-bubble CD significantly lowers the acceleration efficiency of the FS, but increases the injection rate, and leads to the formation of a reflected shock with a speed of approximately $2000\,\mathrm{km\, s^{-1}}$.
Therefore, the FS becomes too weak to provide efficient acceleration but the reflected shock will eventually interact with other structures and will provide several outgoing shocks which can catch up to the FS at later times. The innermost reflected shock speed increases to approximately $\mathrm{3000\,km\, s^{-1}}$ during its propagation towards the ejecta, and it may re-energise the particles during its passage towards the interior of the remnant. This situation is similar to the efficient acceleration at the reverse shock of a very young SNR, for example Cas A \myblue{\citep{1996ApJ...466..866B}}. But, efficient particle acceleration at the reverse shock requires significant magnetic amplification on account of the weak field in the ejecta (\myblue{\citep{2005A&A...429..569E, 2014ApJ...785..130Z}}).  
After 44000 years, the proton spectrum, shown in the third column of Fig. \ref{fig: Figure 5}, and likewise the electron spectrum illustrated in Fig.\ref{fig: Figure 7}, are much softer than a $E^{-2}$ power law. The FS propagates through a dense and cold medium, and so a huge number of low-energy particles are injected at low momenta, but the compression ratio of the FS is around 2.5. Beyond this time, we see no further change in the proton and electron spectra, and our simulation ends after 46000 years.\\ 
With time, the hydrodynamic structure within the wind bubble becomes very complex as a consequence of the many reflected and transmitted shocks. To model the particle acceleration precisely, resolving all the shocks in the FS downstream is desirable but quite impossible to execute. One possible effect of the limited resolution is that highly energetic particles may experience two or more small shocks as one structure with unusually small or large velocity compression, the latter of which may cause small spectral bumps at higher energy.

\subsection{Non-thermal emission spectra}
\label{subsec:3.3}

\begin{figure}[htp]
\centering
\includegraphics[width=9cm, height=9cm]{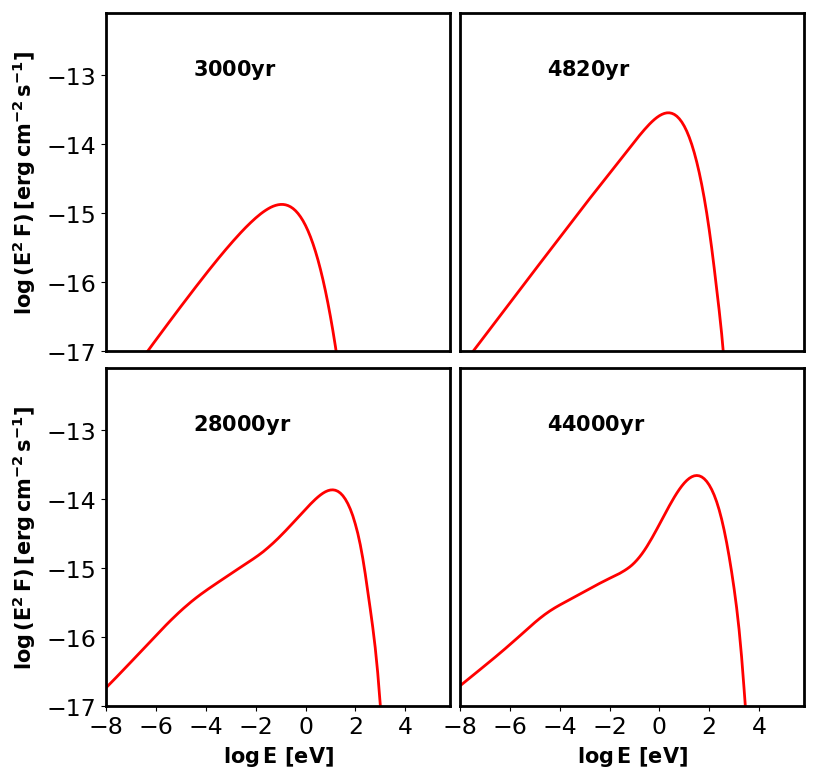}
\caption{\textbf{Spatially-integrated synchrotron (SY) spectra at different ages of the SNR.}}
\label{fig: Figure 8}
\end{figure}

We have considered three non-thermal emission processes: Synchrotron emission, inverse Compton scattering of CMB photons, and the decay of neutral pions. The methods of calculation are described in \myblue{\citet{2013A&A...552A.102T}} and \myblue{\citet{Bhatt_2020}}, respectively. Fig. \ref{fig: Figure 8} and Fig. \ref{fig: Figure 10} depict synchrotron spectra and the gamma-ray spectra, respectively, at four points in time. Fig. \ref{fig: Figure 11} illustrates the energy flux for synchrotron emission and gamma-ray emission during the different evolutionary stages of the remnant. The flux is calculated considering the remnant at $1$~kpc distance. 
\begin{figure}[htp]
\centering
\vspace*{-0.1cm}
\includegraphics[width=9cm, height=6cm]{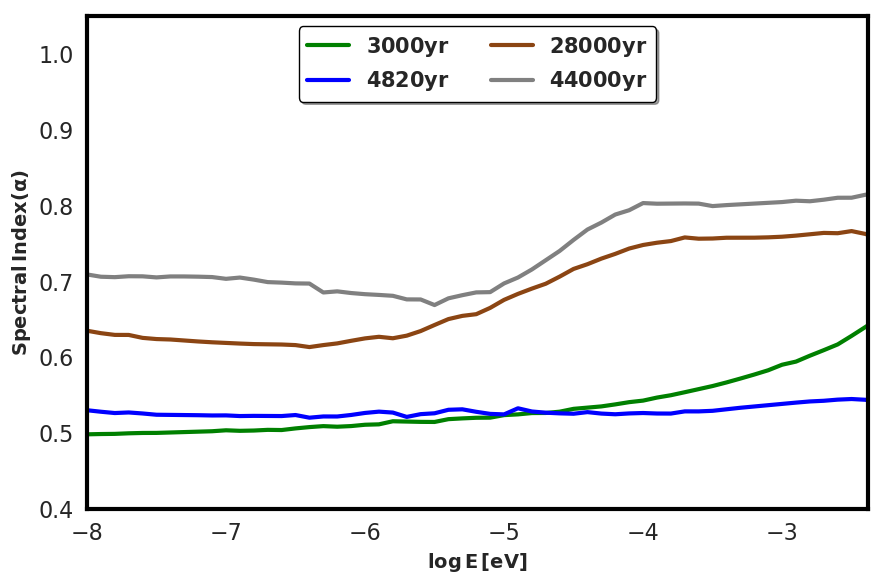}
\caption{\textbf{Variation of the spectral index ($\mathbf{\alpha}$) for synchrotron emission with energy at different ages} where energy flux ($S_{\nu}$) $\propto \nu ^ {-\alpha}$.}
\label{fig: Figure 9}
\end{figure}
The magnetic field strength in our simulation is weak both upstream and downstream of the FS, at least until it reaches the ISM. Therefore, synchrotron cut-off energy above $100\,\mathrm{eV}$ has been achieved only very late in the evolution, but for the first few thousand years, the cut-off energy only reaches near $50\,\mathrm{eV}$. Turbulent magnetic amplification by streaming instabilities or dynamo action have not been considered, although some evidence for that has been observationally obtained \myblue{\citep{1999ICRC....3..468E, 10.1093/mnras/stt1318,2014ApJ...785..130Z}}.
\begin{figure}
\centering
\includegraphics[width=9cm, height=9cm]{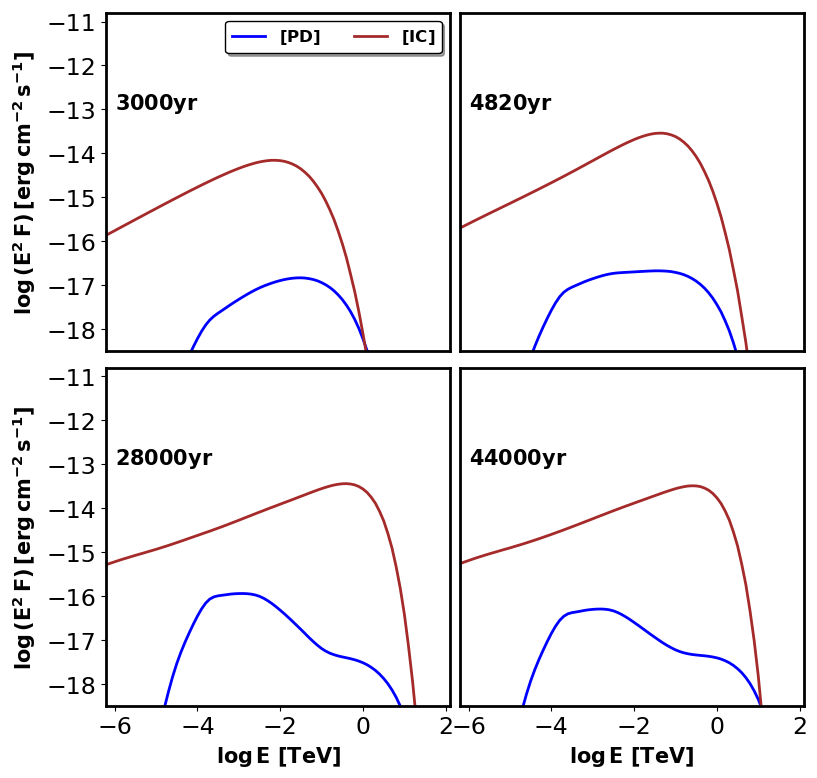}
\caption{\textbf{Spatially-integrated gamma-ray spectra of pion-decay (PD) emission and inverse Compton scattering (IC) at different ages.}}
\label{fig: Figure 10}
\end{figure}
\vspace{5pt}
\par \textbf{FS in the free wind:} In this region, the flux of both the synchrotron and the hadronic emission decrease with time on account of the declining density, $\rho \propto 1/r^2$, and also the weakening magnetic field, $B \propto 1/r$. Panel [a] of Fig. \ref{fig: Figure 11}, region 1 demonstrates that the simulated radio flux at $5\,\mathrm{GHz}$ energy and the X-ray flux in the range $0.1\mathrm{KeV}-10\mathrm{KeV}$ show a power-law decrease with different slopes. It is evident that X-ray emission dominates at the very initial stage of the remnant, but as consequence of the declining maximum achievable energy of electrons, the X-ray emission fades quicker than the radio emission, and radio emission starts to  dominate after around 1100 years. In the gamma-ray flux shown in panel [b] of Fig. \ref{fig: Figure 11}, interestingly inverse Compton emission dominates the high energy (HE) gamma-ray flux except for around initial 60 years, whereas pion-decay emission dominates in the very high energy (VHE) gamma-ray band. This result reflects that initially the remnant expands through dense material and hence, enhanced pion-decay emission has been achieved, and the decreasing electron cut-off energy reduces the VHE inverse Compton flux. At the age of $3000$ years, the synchrotron flux is very low because of the weak magnetic field. Further, the FS propagates through a region of declining density, and consequently the pion-decay emission is also very weak.
\begin{figure}[t]
\includegraphics[width=9cm]{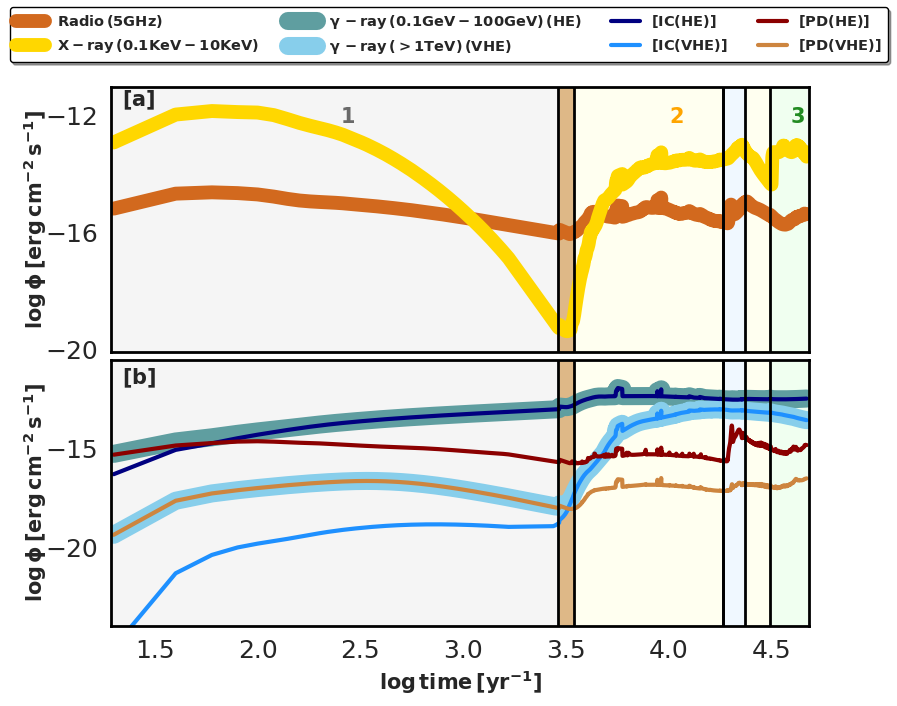}
\caption{\textbf{Evolution of energy flux ($\mathbf{\phi}$) during the lifetime of SNR for synchrotron emission and gamma-ray emission at specific energy ranges.} Region 1, Region 2, and Region 3 denote the free wind, shocked wind and shocked ISM region respectively, distinguished by different colours. The brown shaded region around 3200 years denotes the FS transition from free wind to shocked wind zone, further the blue shaded region around 20000 years denotes the interaction between FS and LBV shell.}
\label{fig: Figure 11}
\end{figure}
\par\textbf{FS in the shocked wind:} As soon as the FS enters this region, the X-ray synchrotron flux starts to grow and eventually dominates over the radio flux on account of the increasing field strength, shown in the region 2 of panel [a] of Fig. \ref{fig: Figure 11}. In this region, both the HE gamma-ray and VHE gamma-ray emissions are dominated by inverse Compton scattering, as the remnant expands inside a region of very low density. The non-thermal emission flux fluctuates as the FS interacts with several discontinuities present in this region.
For example, the slight increase in HE hadronic emission and radio emission near 20000 years depicted in the Fig. \ref{fig: Figure 11} by blue shaded region between 17800 years and 23500 years, indicates an interaction between the FS and the LBV shell. The electron spectra are slightly softened, and so are the synchrotron spectra. Fig. \ref{fig: Figure 9} displays the spectral index of synchrotron emission as a function of photon energy. In the radio band the index is initially around $\alpha\approx 0.53$.
Fig. \ref{fig: Figure 10} indicates that after 4820 years the roughly constant gas density in the shocked wind moderately boosts the pion-decay emission in comparison to that at the later stage of propagation through the free wind region. Inverse Compton emission is in spectral agreement with, but at a lower flux than the observed signal from  RX~J1713.7-3946 \myblue{\citep{2007A&A...464..235A,2015A&A...577A..12F}} and Vela Jr. \myblue{\citep{2018A&A...618A.155S}}. After $28000$ years, the two-component structure of the synchrotron spectrum, that reflects the break in the electron spectrum, becomes visible, and the radio spectral index approaches $\alpha \approx 0.7$.
The inverse-Compton emission now extends to the TeV scale and the pion-decay spectrum is rather soft above a few GeV, but shows a weak bump around a TeV as a consequence of amplified injection in the LBV shell (cf. section \ref{subsec:3.2}).

\par\textbf{FS in the shocked ISM:} After entering into the shocked ISM, the FS propagates in a region with a strong magnetic field, which changes the synchrotron spectra. Our calculated X-ray flux dominates over radio flux in this region also, and although the hadronic HE emission starts to grow in this region, both of the HE and VHE gamma-ray emission have a leptonic origin, illustrated in the Fig. \ref{fig: Figure 11} inside region 3.
We observe very soft spectra from the radio band ($\alpha \approx 0.71$) to the infrared ($\alpha \approx 0.83$). Additionally, the spectral index for pion-decay emission above $10\,\mathrm{GeV}$ reflects the softness of the proton spectra (spectral index $\approx 2.6$). Correspondingly soft gamma-ray has been observed from IC443 and W44, both of which expand in a dense molecular cloud \myblue{\citep{10.1093/mnras/stt1318, 2014A&A...565A..74C}}. The obtained soft radio spectra in our simulation is quite consistent with the data for many Galactic SNRs \myblue{\citep{1977A&A....61...99B, 2009BASI...37...45G, 2014Ap&SS.354..541U, 10.1093/mnras/staa3896}}.

\subsection{Morphology of non-thermal emission}
\label{subsec:3.4}

\begin{figure}
\centering
\includegraphics[width=9cm, height=9cm]{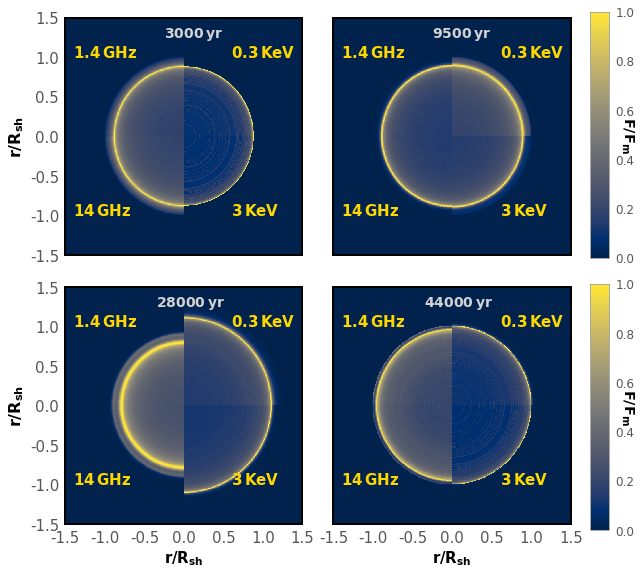}
\caption{\textbf{Normalised intensity maps for synchrotron emission:} Each panel is divided into 4 segments- the left hemisphere is for radio emissions at $1.4$-$\mathrm{GHz}$ in the upper half and at $14$-$\mathrm{GHz}$ in the lower half. The right hemisphere is for the $0.3$-$\mathrm{keV}$ and $3$-$\mathrm{keV}$ X-ray intensity in the upper half and lower half respectively. For each segment, the intensity, $F/F_\mathrm{m}$, is normalised to its peak value, $F_\mathrm{m}$.}
\label{fig: Figure 12}
\end{figure}

\begin{figure}
\centering

\includegraphics[width=9cm, height=9cm]{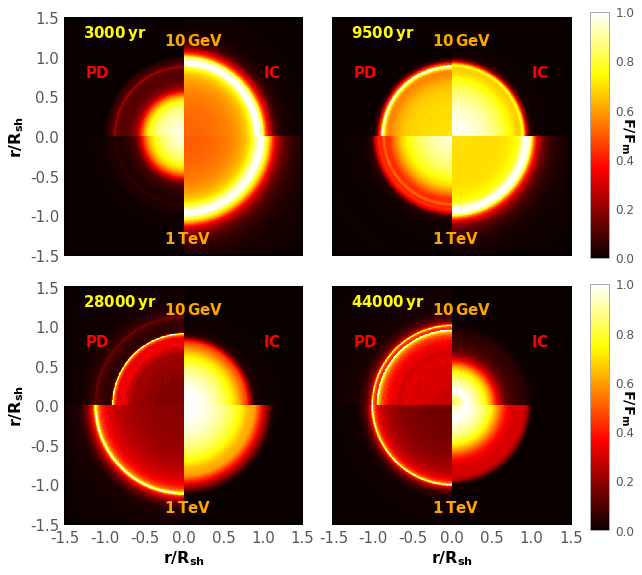}
\caption{\textbf{Normalised intensity maps of pion-decay (PD) and Inverse Compton (IC) emission:} The segments are organised to distinguish the photon energy, $10\,\mathrm{GeV}$ in the upper half and
$1\,\mathrm{TeV}$ in the lower half, and PD on the left hemisphere and IC on the right hemisphere. In each segment, the intensity, $F/F_\mathrm{m}$, is normalised to its peak value, $F_\mathrm{m}$.}
\label{fig: Figure 13}
\end{figure}

We have calculated intensity maps, depicted in Fig. \ref{fig: Figure 12} for synchrotron emission and Fig. \ref{fig: Figure 13} for gamma-ray emissions, respectively, for a fiducial distance of $1000\, \mathrm{parsec}$. To be noted from the figures is the evident variation of the source morphology with the age of the SNR.

The X-ray morphology ($0.3$~$\mathrm{KeV}$ and $3$~$\mathrm{KeV}$) features a thin shell throughout the entire lifetime of the SNR whereas the radio morphology ($1.4$-$\mathrm{GHz}$ and $14$-$\mathrm{GHz}$) shows a comparatively thicker shell and eventually becomes moderately centre-filled when the FS is in the shocked wind, shown at 28000 years in Fig. \ref{fig: Figure 12}. After 3000 years, when the FS propagates through the free stellar wind, the brightest synchrotron emissions in both the radio and the X-ray band emanate from the contact discontinuity between the FS and the RS on account of the strong magnetic field there \myblue{\citep[cf.][]{2004ApJ...609..785L}}. At later stage, when the FS passes through the shocked stellar wind, the synchrotron morphology is essentially the same as at earlier times.
After 28000 years, when the FS is still inside the shocked wind but approaches the CD between the wind bubble and the ISM and already interacted with the LBV shell, the brightest radio emission comes from the region near the contact discontinuity between the FS and the RS as well as the region near the LBV shell whereas the CD of the wind bubble appears X-ray bright as there the magnetic field is twelve times stronger than that immediately downstream of the FS. After 44000 years, when the FS is located in the shocked ISM, the highest radio intensity emanates from a region near the LBV shell whereas the X-ray emission comes from immediately downstream of the FS. The magnetic field downstream of the FS is with a strength below $1\mu\,G$ too weak to produce significant radio emission, but as consequence of diffusion of the electrons in the deep downstream the remnant appears as somewhat centre-filled in the radio band.

In the gamma-ray band, the IC morphology shows a thick shell whereas
the PD emission appears centre-filled at earlier evolutionary stages of the remnant inside the free wind, illustrated in the Fig. \ref{fig: Figure 13} at 3000 years. The maximum IC intensity emanates from the region around the FS for both energies and the interior of the remnant also appears brighter as weak magnetic field downstream of the FS allows electrons to deeply penetrate. PD emission primarily comes from two regions, the dense ejecta in the interior of the SNR and a broad region near the contact discontinuity between the FS and the RS. Later at 9500 years,  the IC intensity at $\mathrm{10\,GeV}$ comes from the entire region downstream of the FS, and at $\mathrm{1\,TeV}$ the IC intensity is the highest in a shell located immediately downstream of the FS. Similarly, the pion-decay emission is seen from almost the entire region interior of the FS, and at $\mathrm{TeV}$ energies most of the flux comes from the SNR ejecta. At 28000 years, the entire region inside the RS is IC bright, specifically where the magnetic field is weak whereas the PD emission at both $10\,\mathrm{GeV}$ and  $1\,\mathrm{TeV}$ feature shell-like structure. The TeV shell is located at the wind bubble CD, while at $10\,\mathrm{GeV}$ the highest PD intensity comes from the region around the LBV shell. Finally, after FS and wind bubble CD interaction, the IC emission appears centre-filled inside the RS, as it did at early times, and PD emission shows a shell-like morphology, depicted in the Fig. \ref{fig: Figure 13} at 44000 years.
\par In reality, the emission maps will be more complex and patchy, because the distinct shell-like morphology reflects the
spherical symmetry in our the 1-D simulations. In particular the Rayleigh-Taylor instability \myblue{\citep{2010A&A...515A.104F}} may break the contact discontinuities in the wind bubble and in the SNR into fragments.

\section{Conclusions}
We have explored the structure of the wind bubble created by a 
$60\,M_{\sun}$ star with solar metallicity formed by the mass-loss of the star from the Zero Age Main Sequence (ZAMS) phase to the pre-supernova stage to study the interactions of the eventual SNR shock with the modified CSM during its passage through the wind bubble. Our simulations of particle acceleration at the forward shock of the SNR suggest that the spectra and flux of energetic particles in core-collapse SNRs are significantly influenced by the structure of the wind bubble. The modification in particle acceleration is thus intertwined to the evolution history of the progenitor star and depends on its properties, such as the ZAMS mass, the metallicity, and the rotation. Our simulations demonstrate the impact of the various discontinuities including the wind-termination shock, a dense LBV shell, and the wind bubble CD, and also the effect of shock-merging on particle acceleration.  

The spectra of accelerated particles depend on the interactions of the FS with the CSM as well as the CSM magnetic-field model. Throughout the propagation of the FS in the hot wind bubble and shocked ISM, beginning at an age of about $3,300$~years, softer particle spectra are persistently observed, on account of a relatively small sonic Mach number of the forward shock. For protons above $10\,\mathrm{GeV}$ energy spectral index reaches around $2.5$. Further, total production spectrum released into the interstellar medium, calculated at 46000 years shows broken power-law with spectral index $s\approx 2.4-2.5$ above $10\,\mathrm{GeV}$ energy. This is broadly consistent with the spectral shape of injection spectrum at higher energy, required by propagation models for the galactic CRs \myblue{\citep{2000ApJ...537..763S, 2007ARNPS..57..285S}}. Besides the small Mach number of the FS, neutral particles in the shocked ISM may also have significant impact on particle acceleration \myblue{\citep{2009ApJ...703L..59O, 2010ApJ...721L..43O}}, but are not considered here, and so is non-linear DSA \myblue{\citep{1981ApJ...248..344D, 1999ApJ...526..385B, 2001RPPh...64..429M}}. 

The spectra and morphology of non-thermal emission reflects the spectral distributions of particles. The gamma-ray emission in our model is dominated by the leptonic contributions, and even that provides a relatively low flux. The pion-decay emission is likely not observable, but has a two-component structure in the spectrum after the interaction of FS with the LBV shell. This feature should be brighter and hence be possibly observable, if the progenitor star sat in a high-density environment. The IC morphology varies between shell-enhanced and centre-filled, whereas the pion-decay emission has a centre-filled to shell-like morphology.
It is challenging to detect an extended object with radius exceeding $80$~pc after 45,000 years ($5^{\circ}$ for a distance of $1\,\mathrm{kpc}$) with a flux as low as we calculate. The flux may be higher for a high-density ISM and for efficient magnetic-field amplification in the remnant, and so there is a possibility to observe with the next generation of observatories, such as SKA, CTA, and LHAASO.  
Although our simulation of an SNR of a progenitor with $60\,M_{\odot}$ ZAMS mass is entirely based on theoretical reasoning, a few remarks about SNR G$150.3+4.5$ (discussed in \myblue{\citet{2020A&A...643A..28D}}) with angular size $\sim 3^{\circ}$ can be offered. The very high shock velocity expected for this extended SNR expanding in a low ambient density suggests a core-collapse scenario with a large wind bubble and the FS expanding in the shocked wind. The predicted maximum cut-off energy for particles ($ 5\,\mathrm{TeV}$) and softer radio spectral index from some regions of this extended SNR are
consistent with the results of our simulation. Additionally, we also find an IC-dominated $\gamma$-ray spectrum as predicted for SNR G$150.3+4.5$.
In conclusion, we have investigated the evolution of an SNR with a Wolf-Rayet progenitor considering Bohm-scaling of diffusion downstream and immediately upstream of the FS. \myblue{\citet{2000A&A...357..283B}} estimated the maximum energy for accelerated protons and the cut-off energy for expected $\gamma$-ray flux to be $10^{14}\,\mathrm{eV}$ and $10^{13}\,\mathrm{eV}$, respectively, considering Bohm diffusion during the expansion of SNR with Wolf-Rayet progenitor and an ejecta mass comparable to that in our simulation. Although our model yields consistent results, the Bohm limit for CR diffusion may be too optimistic. Considering CR streaming instability and Kolmogorov non-linearity in magneto-hydrodynamic waves, \myblue{\citep{2003A&A...403....1P, 2005A&A...429..755P}} estimated analytically that for the ejecta-dominated stage the maximum energy may exceed the "knee" but at the later Sedov phase it can be reduced to $10\,\mathrm{GeV}$. A future study including a diffusion model based on the resonant streaming instability and magnetic field amplification may indicate additional observational signatures. 

\bibliographystyle{aa}
\bibliography{ref}

\begin{appendix}

\section{Effect of shock-shock tail-on interactions on particle spectra}
\label{Appendix A}
When a reflected shock catches up with the forward shock, there is a limited time window of duration $t_i$, in which particles might be able to diffusively cross both shocks and thus probe the full compression ratio of the two-shock system. That compression ratio across both shocks is typically larger than four, and thus a spectral hardening could be expected. However, to have any effect on the spectrum, the interaction time, $t_i$, needs to be longer than the acceleration time of particles. Particles crossing the trailing shock toward the upstream region can reach the leading shock, if it is located within the characteristic distance $L$ that is given by,
\begin{linenomath*}
\begin{align}
    L &= \frac{D_{1\text{,} 2}}{V_\text{sh,2}} \text{ , }
    \label{app:eq1}
\end{align}
\end{linenomath*}
where $D_{1\text{,} 2}$ is the spatial diffusion coefficient of particles between two shocks at a given energy and $V_\mathrm{sh,2}$ is the speed of the trailing shock. It follows that the time to collision is
\begin{linenomath*}
\begin{align}
    t_i &= \frac{D_{1\text{,} 2}}{V_\text{sh,2}}\frac{1}{\Delta v} \text{ , }
\end{align}
\end{linenomath*}
{where $\Delta v$ denotes the difference in the propagation speed between both shocks. It can be related to the speed of the two shocks, 
\begin{linenomath*}
\begin{equation}
    \Delta v = V_\text{sh,2} -\frac{1}{\kappa_1} V_\text{sh,1} \text{,}
\end{equation}
\end{linenomath*}
where $\kappa_1$ is the compression ratio of the leading shock. Assuming the first shock is strong, $\kappa_1=4$, and an adiabatic index $\gamma=5/3$, the sonic Mach number of the second shock is
\begin{linenomath*}
\begin{equation}
    M_2 = \frac{V_\text{sh,2}}{V_\text{sh,1}} 
    \frac{4}{\sqrt{5}}\text{.}
    \label{app:eq4}
\end{equation}
\end{linenomath*}
Having a trailing shock ($M_2 >1$) obviously requires that $\Delta v >  V_\text{sh,1} (\sqrt{5}-1)/4$, and so the time to collision cannot be made arbitrarily long. 
We can now express the interaction time in terms of $V_\text{sh,1} $ and $M_2$,
\begin{linenomath*}
\begin{equation}
   t_i = \frac{D_{1\text{,} 2}}{V_\text{sh,1}^2}\frac{16}{5 M_2^2}
   \frac{1}{1-\frac{4}{\sqrt{5}\kappa_1 M_2}}\simeq 
    \frac{D_{1\text{,} 2}}{V_\text{sh,1}^2}\frac{16}{5 M_2^2}
   \frac{1}{1-\frac{1}{\sqrt{5} M_2}}
\text{.}
\end{equation}
\end{linenomath*}
The total compression at the two-shock system is (again for $\kappa_1=4$)
\begin{linenomath*}
\begin{equation}
   \kappa_\mathrm{tot}=\frac{4}{1-\sqrt{5}\,\frac{3}{4}\,\frac{M_2^2-1}{M_2}}
   \text{,}
   \label{app:eq5}
\end{equation}
\end{linenomath*}
where a negative value implies that the far-downstream flow is faster than the first shock. It is evident that a very moderate Mach number of the trailing shock, $M_2$, is sufficient to significantly raise the total compression ratio, $\kappa_\mathrm{tot}\gg 4$, which would with time lead to very hard particle spectra. The question is whether or not there is sufficient time to establish such a hard spectrum.}

We can extend the analysis of \myblue{\citet{1991MNRAS.251..340D}} to see that the relative momentum gain per shock-acceleration cycle,
\begin{linenomath*}
\begin{equation}
\frac{\Delta p}{p} \approx \frac{4}{3} \frac{\kappa_\mathrm{tot}-1}{\kappa_\mathrm{tot}} \frac{V_\text{sh,1}}{c} \text{,}
\end{equation}
\end{linenomath*}
is only weakly enhanced at the two-shock system, whatever the total compression. The mean residence time downstream of the leading shock is 
\begin{linenomath*}
\begin{equation}
\Delta t_d\approx \frac{4}{c} \int_0^\infty dy\ \exp\left(-\int_0^y dx\ \frac{V_d(x)}{D(x)}\right) \text{,}
\end{equation}
\end{linenomath*}
where in the absence of the trailing shock $V_d=V_\text{sh,1}/\kappa_1$. Downstream of the trailing shock, the flow speed and possibly the diffusion coefficient will change. For constant flow speed and diffusion coefficient between the shocks, and integrating only over the distance between the shocks, $L$, we obtain a strict lower limit to the duration of an acceleration cycle,
\begin{linenomath*}
\begin{equation}
t_c > \frac{16D_{1\text{,} 2}}{cV_\text{sh,1}}\left[1-\exp\left(-\frac{1}{\sqrt{5} M_2}\right)\right] \text{,}
\end {equation}
\end{linenomath*}
where we used equations~\ref{app:eq1} and \ref{app:eq4}.
The acceleration-time, $t_a=t_c\, p/\Delta p$, of a particle is
\begin{linenomath*}
\begin{align}
    t_a &\gtrsim 
    12\frac{D_{1\text{,} 2}}{V_\text{sh,1}^2} 
    \frac{\kappa_\mathrm{tot}}{\kappa_\mathrm{tot}-1} \left[1-\exp\left(-\frac{1}{\sqrt{5} M_2}\right)\right] \nonumber \\
     &\simeq \frac{12}{\sqrt{5} M_2}\frac{D_{1\text{,}2}}{V_\text{sh,1}^2} 
    \frac{\kappa_\mathrm{tot}}{\kappa_\mathrm{tot}-1}\text{.}
\end{align}
\end{linenomath*}
Comparison with eq.~\ref{app:eq5} yields the ratio of the relevant time-scales is given by,
\begin{linenomath*}
\begin{equation}
    \frac{t_i}{t_a} \lesssim \frac{4\,(\kappa_\mathrm{tot}-1)}{3\, \kappa_\mathrm{tot}}\,\frac{1}{\sqrt{5} M_2-1} \text{.}\label{eq:tita}
\end{equation}
\end{linenomath*}
Given that we ignored time spent upstream of the leading shock or downstream of the trailing shock, we can conclude that particles can see the full compression of both shocks combined for less than a single acceleration time. A similar result is found for the post-collision phase, when the two shocks move in opposite direction and separate very quickly, as well as for head-on collisions. All in all, tail-on shock collisions can not produce significant spectral features.

\section{Animation of the evolution of FS inside the wind bubble}
\label{Appendix B}

\begin{figure*}
\movie[externalviewer]{\includegraphics[width=18.3cm]{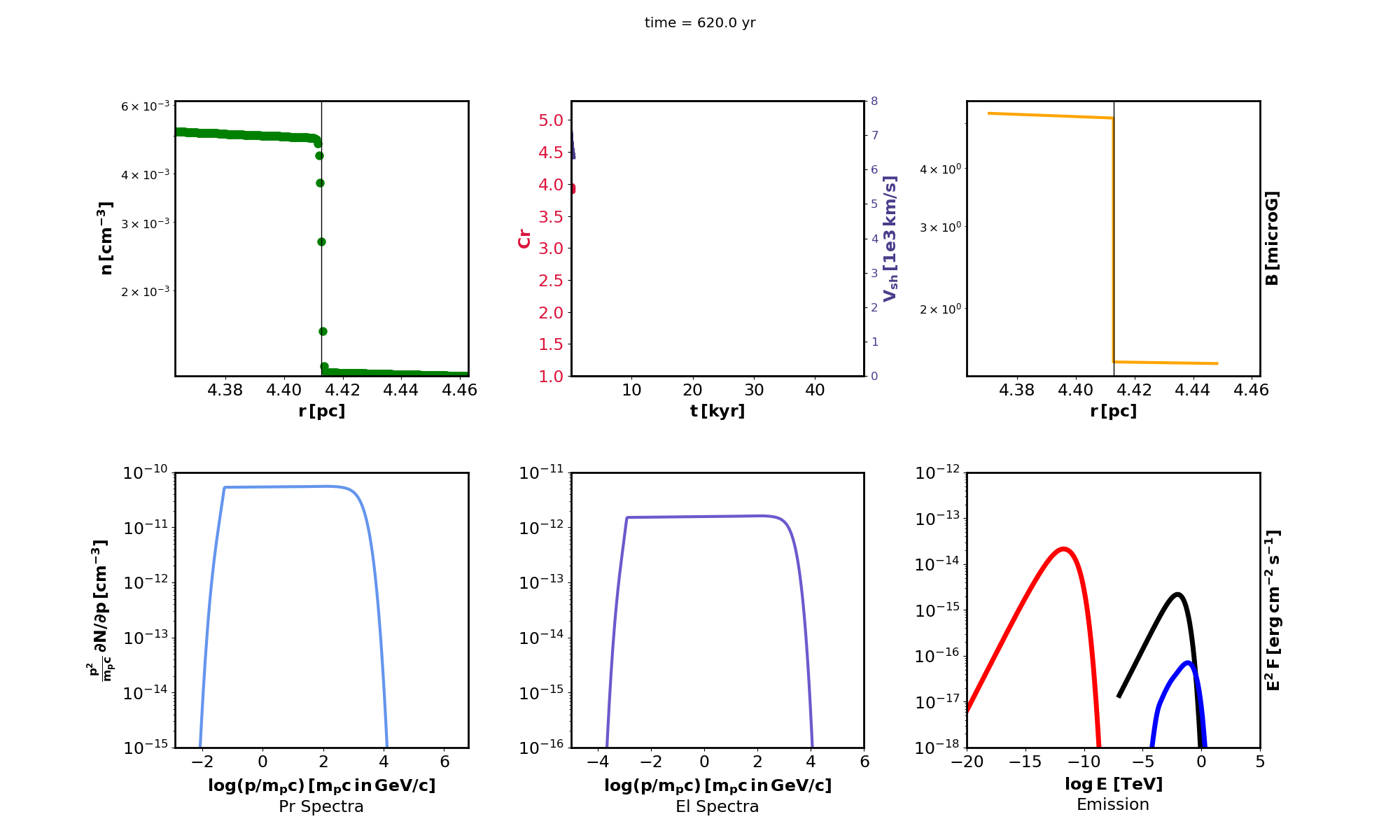}}{emissions_animation.mp4}
\caption{\textbf{Evolution of forward shock.}  (i) Upper row:  The left panel illustrates the gas density, $\mathrm{n}$, as a function of radius, the second panel shows the compression ratio, $\mathrm{Cr}$, and the speed of the sub-shock, $\mathrm{V_{sh}}$, as functions of time, and the third panels depicts the magnetic-field strength near the sub-shock, $\mathrm{B}$. The black vertical line denotes the position of the SNR forward shock.
(ii) Lower row: The first panel shows proton (Pr) spectra volume-averaged downstream of the FS. The corresponding electron (El) spectra are illustrated in the second panel. The third panel shows the spectra of synchrotron emission (red), Inverse Compton emission (black) and pion-decay emission (blue).}
\label{fig: Fig. A.1}
\end{figure*}

\myblue{Fig \ref{fig: Fig. A.1}} includes flow number density($n$) and magnetic field configuration (B) in the vicinity of forward shock along with the sub-shock compression ratio (Cr), shock velocity ($V_\mathrm{sh}$), and proton, electron and non-thermal emission spectra for the entire time span of the simulation at different time steps.

\end{appendix}

\end{document}